\documentclass[twoside,twocolumn,english,aps,prb,longbibliography]{revtex4-2}
\usepackage[T1]{fontenc}
\usepackage[latin9]{inputenc}
\usepackage{amsmath}
\usepackage{amssymb}
\usepackage{graphicx}

\makeatletter
\@ifundefined{textcolor}{}
{%
 \definecolor{BLACK}{gray}{0}
 \definecolor{WHITE}{gray}{1}
 \definecolor{RED}{rgb}{1,0,0}
 \definecolor{GREEN}{rgb}{0,1,0}
 \definecolor{BLUE}{rgb}{0,0,1}
 \definecolor{CYAN}{cmyk}{1,0,0,0}
 \definecolor{MAGENTA}{cmyk}{0,1,0,0}
 \definecolor{YELLOW}{cmyk}{0,0,1,0}
}

\makeatother

\usepackage{babel}

\begin{document}
\title{Quantum Frustration as a Protection Mechanism in Non-Topological Majorana
Qubits}
\author{E. Novais}
\affiliation{Centro de Ci\^{e}ncias Naturais e Humanas, Federal University of ABC,
Brazil}
\date{\today{}}
\begin{abstract}
I analyze the decoherence of a $\pi$-junction qubit encoded by two
co-located Majorana modes. Although not topologically protected, the
qubit leverages distinct spatial profiles to couple to two independent
environmental baths, realizing the phenomenon of quantum frustration. This
mechanism is tested against the threat of quasiparticle poisoning
(QP). I show that frustration is effective against Ohmic noise ($s=1$)
and has some protection for $0.76<s<1$ sub-Ohmic noise. However,
the experimentally prevalent $1/f$ noise ($s\to0$) falls deep within
the model's localized phase, where frustration is insufficient. This
causes spontaneous symmetry breaking and catastrophic decoherence.
The qubit's viability depends on what the effective environment is
that these local Majorana wave functions experience.
\end{abstract}
\maketitle

\section{introduction}

Decoherence and dissipation have long represented major obstacles
to the realization of quantum computing. These processes, which describe
the loss of quantum coherence through interaction with the environment,
rapidly destroy the delicate superposition and entanglement that quantum
devices rely upon. The persistent challenge of protecting quantum
information has therefore motivated an intensive search for intrinsically
robust quantum systems, a pursuit that has brought topological concepts
to the forefront of modern physics.

The central idea underlying topological protection is that quantum
states belonging to distinct topological sectors cannot be connected
by any local perturbation. In other words, local noise or imperfections
in material structure cannot induce transitions between these sectors.
Qubits encoded in such states are thus inherently resilient, offering
a pathway toward fault-tolerant quantum computation~\citep{freedman2002topologicala}.

Before topology emerged as the dominant paradigm in the quest to overcome
decoherence, several hardware-level strategies were developed to protect
quantum information. These include decoherence-free subspaces and
dynamical decoupling, both of which suppress environmental coupling
through symmetry or time-dependent control. Closely related to these
mechanisms is the concept of quantum frustration of decoherence~\citep{castroneto2003quantum,novais2005frustration}.
Decoherence may be understood as the environment\textquoteright s
selection of a pointer basis in the system. However, if a system couples
simultaneously to two independent environments whose pointer bases
are incompatible, their competing influences can strongly suppress
overall decoherence.

This manuscript explores the intersection between topological protection
and quantum frustration. Specifically, I show that a qubit defined
by two Majorana fermions at a $\pi$-junction~\citep{lutchyn2010majorana,oreg2010helical,klinovaja2012transition,alicea2012new}
experiences frustration of decoherence. Because each Majorana mode
possesses a distinct spatial profile, the two couple to effectively
independent environments. As a result, even though this qubit is not
topological in the strict sense, it acquires a remarkable degree of
protection against environmental noise. 

In the first part of the manuscript I explore three possible experimental
setups to build the qubit in a semiconductor quantum wire. After this
initial discussion the manuscript focuses on how trapped charges and
electromagnetic fluctuations affect the qubit. On one hand, if there
are only a few of these charges, the noise model corresponds to the
usual Random Telegraph Noise of mesoscopic quantum devices. On the
other hand, if there is a dense set of trapped charges the effective
noise model is a generalization of the spin-boson model. In this situation,
the noise is characterized by the power noise spectrum,~$S\left(\omega\right)\sim\omega^{s-1}$.
The particular value of $s$ is an open experimental question, since
the qubit has a local spatial profile. The main result of the manuscript
is that quantum frustration of decoherence protects the qubit against
Ohmic noise, $s=1$, and provides partial protection for sub-Ohmic
noise with $0.76<s<1$.

This manuscript is organized as follows. In Sec.~\ref{sec:From-the-Kitaev}
I review the standard Kitaev chain model and some of its experimental
proposals. In Sec.~\ref{sec:the-qubit-in} I introduce the $\pi$-junction
qubit and discuss its possible physical implementations. In sec.~\ref{sec:The-noise-model}
the possible noise mechanisms are presented and the effective noise
model for a dense set of trapped charges is argued. In Sec.~\ref{sec:decoherence}
the decoherence in the spin-boson model is reviewed and the quantum
frustration effects are discussed. Finally, in Sec.~\ref{sec:Dicussion-and-conclusions}
some final remarks and conclusion are presented.

\section{From the Kitaev model to real quantum wires\label{sec:From-the-Kitaev}}

A transformative step to protect quantum information was taken by
A. Kitaev in his seminal 2001 paper~\citep{kitaev2001unpaired}, which
reinterpreted the anisotropic XY model of Lieb, Schulz, and Mattis~\citep{lieb1961two}
through the lens of topology. Kitaev\textquoteright s proposal demonstrated
that a one-dimensional p-wave superconducting chain hosts Majorana
edge modes~\citep{read2000paired} that can serve as topologically
protected qubits. Because these modes carry no electric charge, they
do not couple to charge fluctuations, effectively shielding them from
one of the primary sources of decoherence. This conceptual breakthrough
ignited a vast field of research~\citep{lutchyn2010majorana,oreg2010helical,gangadharaiah2011majorana,alicea2012new,aghaee2023inasal,klinovaja2012transition,prada2020andreev}
and laid the foundation for one of the leading solid-state quantum
computing platforms~\citep{microsoftazurequantum2025interferometric,aghaee2025distinct}.
Although the experimental confirmation of Majorana-based qubits remains
an open, and sometimes contentious, issue~\citep{prada2020andreev,legg2025comment},
the theoretical impact of Kitaev\textquoteright s insight is unquestionable.

Experimentally confirming the existence of Majorana fermions in condensed-matter
systems has proven to be a formidable challenge. The same properties
that make these quasiparticles so attractive for quantum information
processing, that is their insensitivity to local perturbations,
also make them extremely difficult to detect directly. Most available
signatures are therefore indirect and, consequently, open to competing
interpretations. In fact, it has been argued that the topological
superconducting gap required for Majorana protection might not even
form in the semiconductor quantum wires most frequently used for these
experiments~\citep{legg2025comment}. From this perspective, the signals
reported by Microsoft in Ref.~\citep{microsoftazurequantum2025interferometric}
could plausibly originate from disorder effects rather than from genuine
Majorana modes~\citep{legg2025comment}.

The Kitaev model~\citep{kitaev2001unpaired} provides one of the simplest
theoretical realizations of a topological superconductor. It describes
a chain of spinless fermions with a pairing term,

\begin{equation}
H_{K}=-\sum_{n=0}^{N-2}\left(c_{n}^{\dagger}c_{n+1}+\gamma e^{i\varphi}c_{n}^{\dagger}c_{n+1}^{\dagger}+h.c.\right)-\mu\sum_{n=0}^{N-1}c_{n}^{\dagger}c_{n},\label{eq:Kitaev}
\end{equation}
where the nearest-neighbor hopping amplitude was set to $1$.

The standard approach to solve this model~\citep{lieb1961two} is to
define the Majorana modes,
\begin{align}
\eta_{n} & =c_{n}+c_{n}^{\dagger},\label{eq:eta}\\
\nu_{n} & =-i\left(c_{n}-c_{n}^{\dagger}\right),\label{eq:nu}
\end{align}
where $\left\{ \eta_{n},\eta_{m}\right\} =\left\{ \nu_{n},\nu_{m}\right\} =2\delta_{nm}$,
$\left\{ \eta_{n},\nu_{m}\right\} =0$. The single particle spectrum that
emerges is characterized by the superconducting gap,  and lies within the
topological phase for $\left|\mu\right|<2$, where there are exponentially localized Majorana states with zero-energy at the boundaries. 
If the Fermi level is above the energy gap, the parity of these Majorana states cannot fluctuate. In this manuscript I will consider the case where the Fermi level is within the gap.

In general, a quadratic
model like Eq.~(\ref{eq:Kitaev}) can be written as
\begin{align}
H_{q} & =\frac{i}{2}\sum_{n=-\frac{N}{2}}^{\frac{N}{2}-1}\left[\alpha_{n}\eta_{n}\nu_{n+1}-\beta_{n}\nu_{n}\eta_{n+1}+\mu\eta_{n}\nu_{n}\right]+\frac{\mu N}{2},\label{eq:majorana-model}
\end{align}
where $\alpha_{n}=t_{n}+\gamma_{n}$, $\beta_{n}=t_{n}-\gamma_{n}$,
$t_{n}$ is the hopping amplitude and $\gamma_{n}$ the superconducting
pairing between sites $n$ and $n+1$. A short review on how to find
the analytical solutions is given in Appendix~\ref{sec:diagonalizing-the-free}.

Although deceptively simple in appearance, this Hamiltonian is difficult
to reproduce in a real material. The most widely explored route employs
semiconducting nanowires with strong Rashba spin--orbit coupling,
such as $InAs$ and $InSb$, which offer large g-factors ($\approx15$
and $\approx50$ in bulk, respectively) and pronounced spin--orbit
interactions~\citep{lutchyn2010majorana,oreg2010helical,klinovaja2012transition,alicea2012new}.
A sketch of this mapping is presented in Appendix~\ref{sec:Semiconductor-quantum-wires}.

A central point in the mapping between the semiconducting nanowire
and the Kitaev chain is that the effective pairing amplitude $\gamma e^{i\varphi}$
, is not a fixed parameter, but rather depends on several intertwined
physical quantities. Specifically, it is determined by the superconducting
phase of $\Delta$, the direction of the external magnetic field $B_{0}$
and \emph{crucially} the spin-orbit vector $\hat{e}$ (see Ref. ~\citep{alicea2011nonabelian}
for the details),

\begin{equation}
\gamma e^{i\varphi}=\frac{\alpha\Delta e^{i\arg\left(\hat{e}\right)}}{g\mu_{B}\left|B_{0}\right|},\label{eq:effectivve_pairing}
\end{equation}
where $\alpha$ is the strength of the spin-orbit coupling.

Although every step of the derivation is physically plausible, it
is not yet evident that existing experiments unambiguously confirm
the detection of Majorana fermions. 

\section{the qubit in a $\pi$-junction\label{sec:the-qubit-in}}

The name $\pi$-junction is derived from the usual discussion of Josephson
junctions. In a junction the phase difference between the superconducting
order parameters depends sensitively on the properties of the material
forming the layer. A particularly interesting situation arises when
the barrier separating two $s$-wave superconductors is ferromagnetic.
As electrons in a Cooper pair tunnel through the ferromagnetic layer,
they experience opposite exchange fields, which introduce a relative
phase shift between them. When the thickness of the ferromagnetic
layer is appropriately chosen, this process leads to a total phase
difference of $\pi$ between the two superconductors, thereby locking
the relative phase of their condensates at $\pi$. Experimental realizations
of such $\pi$-junction have been reported in $NbN/CuNiNbN$ heterostructures,
where a ferromagnetic barrier of approximately $10\sim20nm$ produces
a $\pi$ phase shift at temperatures around $4K$~\citep{frolov2004measurement,yamashita2017nbnbased}.
A recent approach to create Majorana modes in $\pi$-junctions avoids
the usual route to have a semiconductor nanowire~\citep{pientka2017topological}.
I will follow the more traditional approach and consider the physical
realization of the Kitaev model as described in the previous section.

\begin{figure}
\begin{centering}
\includegraphics[width=0.75\columnwidth]{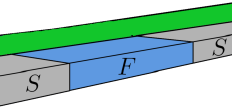}
\par\end{centering}
\caption{\label{fig:pi-1}Two $s$-wave superconductors (gray) with a ferromagnetic
layer between them (blue). On top is the semiconductor wire (green).
The $\pi$-junction in the superconductors induces a $\pi$-junction
in the wire by the proximity effect.}

\end{figure}

\begin{figure}
\begin{centering}
\includegraphics[width=0.95\columnwidth]{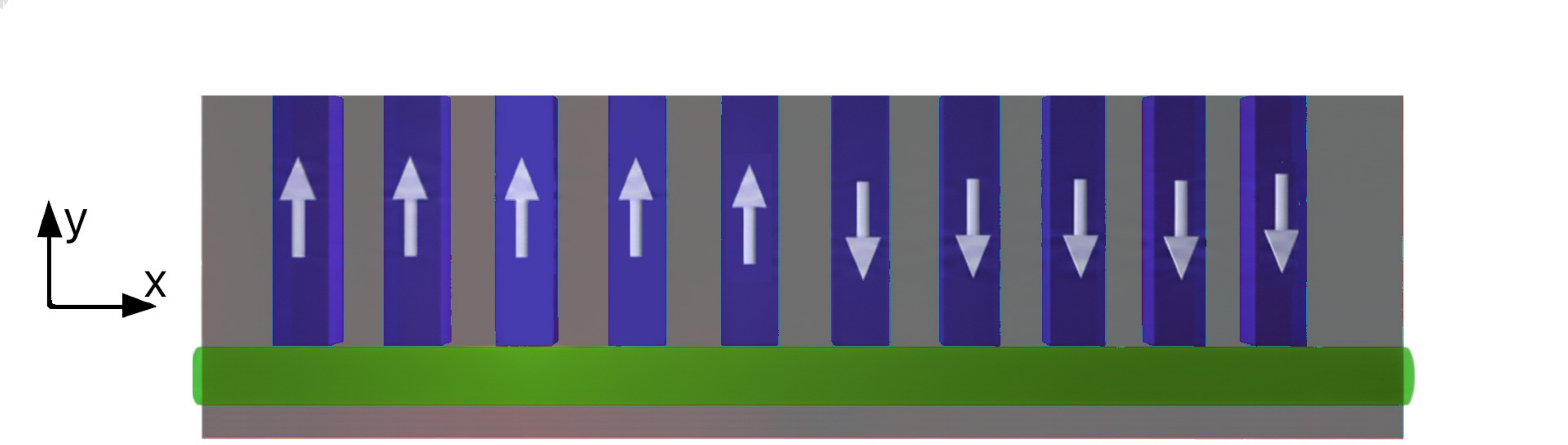}
\par\end{centering}
\caption{\label{fig:semiconductor-wire}Semiconductor wire (green) on top of
an $s$-wave superconductor (gray) with nano-ferromagnets (blue) producing
a magnetic domain wall. Figure adapted from Ref.~\citep{gangadharaiah2011majorana}.}
\end{figure}

My discussion starts exactly as Refs.~\citep{lutchyn2010majorana,oreg2010helical,klinovaja2012transition,alicea2012new}
and assume that $InSb$ or $InAs$ wires are used to construct the
Kitaev chain. Using Eq.~(\ref{eq:effectivve_pairing}), one can identify
three possible mechanisms to realize a $\pi$-junction in the semiconductor
wire.
\begin{enumerate}
\item The most direct scenario occurs when the quantum wire is placed atop
of a superconducting $\pi$-junction. In this configuration, each
side of the wire acquires an effective pairing amplitude from a different
$s$-wave superconductor with a relative phase of $\pi$, see Fig.~\ref{fig:pi-1}.
The physics of this model was described at Ref.~\citep{baldo2023zerofrequency}.
\item A second mechanism involves reversing the direction of the external
magnetic field responsible for the Zeeman splitting, Eq. (\ref{eq:H_z}).
This approach is, in principle, achievable in $InSb$ wires by using
nanopatterned ferromagnets, as proposed in Ref.~\citep{klinovaja2012transition},
see Fig.~\ref{fig:semiconductor-wire}. 
\item The third, and most experimentally practical mechanism, is to reverse the
orientation of the spin-orbit coupling vector, $\hat{e}$ sproposed
on Ref.~\citep{klinovaja2015fermionic}. If a
wire is grown along a given crystallographic direction defining $\hat{e}$
and then cleaved and reoriented, the two segments have opposite spin-orbit
orientations. The junction between these two regions thus realizes
a $\pi$-domain wall in the effective pairing phase, providing a simple
and experimentally accessible route to a $\pi$-junction device. 
\end{enumerate}

\begin{figure}
\begin{centering}
\includegraphics[width=0.75\columnwidth]{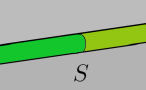}
\par\end{centering}
\caption{\label{fig:invert-spin-orbit}A semiconductor wire is cleaved and
one of the segments is reoriented; the two different green tones represent
the two wires creating a junction. Both are placed on top of an $s$-wave
superconductor (gray).}
\end{figure}

The Cooper pair coherence length defines another important length-scale.
If the separation between the superconductor wires is larger than
this distance, then the junction is defined as long. Conversely, if
the length is smaller it is defined as a short junction~\citep{baldo2023zerofrequency}.

All these scenarios can be described by Kitaev like minimal models.
The analytical solutions presented in Appendix~\ref{sec:diagonalizing-the-free}
show that there are four zero energy modes: 
\begin{enumerate}
\item two Majoranas at the edges of the chain, that I label as $\zeta_{-\frac{N}{2}}$
and $\zeta_{\frac{N}{2}-1}$,
\item and two Majoranas at the junction, that I label as $\zeta_{s}$ and
$\zeta_{a}$.
\end{enumerate}
A Majorana at the end of the wire have a exponentially small overlap
with all the others, hence in the thermodynamical limit it cannot
be combined into a fermion. This is the standard Kitaev argument and
it still holds in these wires. Of course, in reality there is no thermodynamic
limit and a small energy splitting exists. It has been shown that
in more realistic models the electromagnetic environment changes the
topological phase boundary and this energy splitting~\citep{vuik2016effects}. 

The fact that the two Majoranas at the junction are always at zero
energy is much more interesting. The standard Kitaev's argument does
not hold in this case. The deep physical reason that $\zeta_{a,s}$
are insensitive to changes in the chemical potential is the fact that
the superconducting order parameter must go to zero at the center
of the junction.
This is a direct manifestation of the Particle-Hole Symmetry (PHS) inherent to the superconducting state. In this system, the superconducting pairing, which acts as the topological mass, changes sign across the junction. According to the Index Theorem, such a sign change across a domain wall forces the superconducting gap to close, trapping states at the Fermi level. Because these states are their own particle-hole partners, they are initially constrained to remain at $E=0$.

It is important to note that while these Majoranas do hybridize (as they are not "topologically protected" via spatial separation), the resulting fermionic level remains at zero energy due to the phase difference across the junction (see ~\citep[eq. (6)]{alicea2011nonabelian}).
The underlying reason that in a $\pi$ junction the hybridization matrix element is exactly zero, ~\citep[eq. (6)]{alicea2011nonabelian}, is ultimately chiral symmetry. 
In a uniform one-dimensional topological superconductor, the absolute phase can be removed by a global gauge transformation, allowing the Hamiltonian to anticommute with a chiral operator ~\citep{altland1997nonstandard,schnyder2008classification}. In a junction geometry, however, the relative phase difference $\delta\phi$ across the junction is gauge-invariant. Crucially, chiral symmetry is strictly preserved at the special phase differences $\delta\phi = 0$ and $\delta\phi = \pi$. In a $\pi$-junction, the sign change of the pairing potential across the interface forces the two localized Majorana modes to carry the same chirality. Because any finite-energy eigenstate of a chiral-symmetric Hamiltonian must be formed by components of opposite chirality, these two same-chirality Majoranas are strictly forbidden from hybridizing ~\citep{ikegaya2015anomalous,ikegaya2018symmetry}. Because non-magnetic disorder preserves this chiral symmetry, this zero-energy pinning remains even in dirty junctions \citep{asano2006anomalous}. Consequently, the resulting fermionic level remains pinned at exactly zero energy. This feature can be directly observed
in the simulation of real quantum wires~\citep{baldo2023zerofrequency} and not
a consequence of the Kitaev model.

A useful heuristic is to view the
p-wave superconducting wire as the cross section of a two dimensional
p-wave superconductor. In this analogy the $\pi$-junction corresponds
to a vortex in the 2D superconductor and the two Majoranas are then
pinned inside the vortex~\citep{moore1991nonabelions}.
Unlike their boundary counterparts, the Majorana states at the junction
are bound to zero energy by the form of the wave function and are
immune to the electromagnetic environment. 

In a physical semiconductor nanowire, various static perturbations can explicitly break this chiral symmetry. For instance, an external magnetic field that is not perfectly perpendicular to the spin-orbit vector will cause the localized modes to hybridize~\citep{lutchyn2010majorana}. Similarly, a geometric misalignment of the spin-orbit vectors at the junction acts as another symmetry-breaking perturbation~\citep{klinovaja2015fermionic}. As a result of these physical imperfections, the zero-energy crossing is shifted away from a phase difference of exactly $\pi$. When detuned from this exact crossing, the hybridized modes form a finite-energy Fermionic Bound State (FBS) localized at the junction~\citep{klinovaja2015fermionic}.

However, from an experimental standpoint, this shift merely alters the location of the zero-energy states within the experimental parameter space. The direction of the magnetic field and the physical phase difference between the superconductors are tunable parameters. By actively tuning this parameter space to locate the shifted zero-energy crossing~\citep{lutchyn2010majorana}, the experimentalist effectively realizes the Kitaev chain in the $\pi$-phase, thereby establishing the zero-energy states protected by the aforementioned symmetries. Later in the manuscript, I will discuss how fluctuations around the tuning point change these results.

In the remainder of the manuscript I will discuss the decoherence
of the qubit defined by the Majoranas at the $\pi$-junction,

\begin{equation}
d=\zeta_{s}+i\zeta_{a}.\label{eq:Bog-fermion}
\end{equation}
These two Majorana modes, $\zeta_{s,a}$, form a degenerate zero-energy
state. The two logical states of the qubit, $\left|0\right\rangle $ and
$\left|1\right\rangle $, are the two fermion parity eigenstates of
this zero-energy level. 

This is explicitly not a topologically protected qubit. Since both
Majoranas are co-located at the junction, they are \emph{local} and
can hybridize with any nearby fermionic level. However, a crucial
property of this qubit is that its constituent Majoranas, $\zeta_{a}$
and $\zeta_{s}$, have distinct spatial profiles (e.g., symmetric
and antisymmetric). This difference will be the central point of the
noise model in the next section.

\section{The noise model\label{sec:The-noise-model}}

A naive look at the $\pi$-junction qubit is misleadingly optimistic.
One might consider two common noise sources and conclude the qubit
is robust:
\begin{enumerate}
\item Electromagnetic (Charge) Noise: The qubit is a zero-energy state.
Based on its origin as a node in the wave function, its existence
is insensitive to static, global shifts in the chemical potential
$\mu$. This suggests the qubit should be immune to low-frequency
electromagnetic fluctuations, which is effectively a fluctuating chemical
potential. As can be observed in the analytical solution on Appendix~\ref{sec:diagonalizing-the-free}
the spatial profile of the Majoranas do depend on the chemical potential,
and this is going to play a significant role in the discussion below. 
\item Discrete Andreev Bound States (ABS): The junction may host a discrete
set of trivial ABS~\citep{sauls2018andreev,lidal2025andreev}. If one
of these states is accidentally \textquotedbl in resonance\textquotedbl{}
(at zero energy), it would strongly couple to the qubit. However,
as argued in Ref.~\citep{baldo2023zerofrequency}, these states are
\textquotedbl manageable.\textquotedbl{} Their energy is highly sensitive
to $\mu$, so a gate voltage can be used to \emph{detune} them and
move them away from resonance.
\end{enumerate}
These two considerations suggest a highly robust qubit. This picture,
however, neglects the most dominant decoherence channels: \emph{quasiparticle
poisoning }(QP) from a \emph{soft gap}~\citep{glazman2021bogoliubov}
and phase fluctuation of the $s$-wave superconductors.

\subsection{Quasiparticle Poisoning from a Soft Gap}

The ideal superconductor that I considered so far has a \emph{hard}
energy gap, meaning there is a vanishing density of states (DOS) inside
the gap. Real superconductors, particularly hybrid systems, invariably
contain disorder\cite{ahmed2025oddfrequency,ahmed2025anomalous}. This disorder creates a large number of localized
Andreev-like states and together with a finite lifetime to the Cooper
pairs due to phonons and impurities, creates a quasi-continuum set
of in-gap levels. This effect, known as a \emph{soft gap}, results
in a finite, non-zero DOS at all energies, including $E\approx0$~\citep{dynes1978direct,glazman2021bogoliubov}.

This \textquotedbl soft gap,\textquotedbl{} or Dynes continuum~\citep{dynes1978direct},
provides a reservoir of quasiparticles that can tunnel into the qubit
and flip its parity. This is quasiparticle poisoning and it is known
to plague the traditional Majorana qubits in the topological Majorana
qubits in hybrid nanowire platforms~\citep{rainis2012majorana}. Therefore,
the local fermionic levels that hybridize with the qubit are not just
a few discrete, manageable ABS, but potentially the entire continuum
of \emph{Dynes states}.

An intuitive microscopic model would be

\begin{equation}
H_{\text{low energy}}=\sum_{k}\varepsilon_{k}f_{k}^{\dagger}f_{k}+\lambda_{k}f_{k}^{\dagger}d+\lambda_{k}d^{\dagger}f_{k},\label{eq:low_energy_1}
\end{equation}
where $f_{i}$ are standard fermionic operators, and $\lambda_{k}$
is the hybridization, that can always be made real by a gauge transformation
on $f_{k}$ and $\varepsilon_{k}$ are energies within the superconducting
gap.

Although Eq.~(\ref{eq:low_energy_1}) is very standard in discussing
the Majorana modes and local levels, it is not entirely correct. The
two Majoranas, $\zeta_{a,s}$, that compose the $d$ fermion have
distinct spatial profiles. Hence, it is not correct to describe the
hybridization between these fermionic levels as a simple parameters,
$\lambda_{k}$.

Consider the following example. Imagine a local state exactly at
the middle of a short junction at the surface of the $s$-wave superconductor.
By symmetry it will hybridize with the original fermionic operators
following the symmetric Hamiltonian

\begin{equation}
H_{\text{example}}=\sum_{i=0}^{\frac{N}{2}}\lambda_{i}\left(c_{i}^{\dagger}+c_{-i}^{\dagger}\right)f+\text{h.c.~.}
\end{equation}
In a short junction $\zeta_{s}$ is a symmetric function of the $c_{i}$'s
(peaked at the center of the junction), while $\zeta_{a}$ is anti-symmetric
(zero at the center). Therefore, the symmetric $f$ state of this
example will couple only to $\zeta_{s}$. In the case of a long junction,
where the $\zeta_{a,s}$ are spatially located at the junctions walls
this is even more clear.

This implies that  each
Majorana partner experiences its own, independent decoherence environment.
A general discussion for the reasons behind this conclusion can be found in
appendix~\ref{sec:independent-baths}. However, it can be understood
through a scattering channel analogy. Within the narrow low-energy window of
the "soft gap," it is a standard phenomenological assumption that the  environment
has no spatial structure. Following standard scattering theory, the
an environment now described as a continuum of incoming and outgoing going
waves can be
can be expanded into distinct scattering channels based on parity.
The symmetric Majorana mode $\zeta_s$, possessing an antinode at the junction center, acts as an $s$-wave scatterer and couples exclusively to the even-parity channel of the bath. Conversely, the antisymmetric mode $\zeta_a$, possessing a node at the center, acts as a $p$-wave scatterer and couples exclusively to the odd-parity channel.
Because the even and odd scattering channels of the free-fermion environment are orthogonal by symmetry, they constitute completely independent noise channels.

The correct Hamiltonian must take the form

\begin{align}
H_{\text{QP}} & =\sum_{i\in\text{Env}_{1}}\varepsilon_{i}f_{i}^{\dagger}f_{i}+\lambda_{i}\left(f_{i}^{\dagger}-f_{i}\right)\zeta_{s}\nonumber \\
 & +\sum_{j\in\text{Env}_{2}}\varepsilon_{j}f_{j}^{\dagger}f_{j}+i\lambda_{j}\left(f_{j}^{\dagger}+f_{j}\right)\zeta_{a}\label{eq:H_QP}
\end{align}
where Env.~1 and Env.~2 are the two different spatial sets of Dynes
states that couple to $\zeta_{s}$ and $\zeta_{a}$, respectively (see appendix~\ref{sec:independent-baths}).

The Majorana-Majorana interaction of Eq.~(\ref{eq:H_QP}) is precisely
what is expected in more realistic models. For instance, it has been
shown that a Majorana at the end of a Kitaev chain interacts with
a small quantum dot precisely through this form~\citep{ruiz-tijerina2015interaction,silva2020robustness}. 

\subsection{Phase Fluctuations and Charge Noise Revisited}

The \emph{immunity} to electromagnetic fluctuations is an oversimplification.
The spatial profiles of the Majoranas do depend on $\mu$. Therefore,
a fluctuating chemical potential, $\mu\left(t\right)$, i.e., $1/f$
charge noise, will cause the wavefunctions to fluctuate. This
fluctuation modulates the coupling parameters $\lambda_{i}(t)$
and $\lambda_{j}(t)$ in Eq.~(\ref{eq:H_QP}), creating a form of
\textquotedbl parameter noise\textquotedbl{} that also adds to dephasing.

A final source of decoherence is phase fluctuations. The $\pi$-junction
requires the effective phase between the two topological segments
to be precisely $\pi$. Any \emph{jitter} $\delta\phi(t)$ away from
this condition introduces a direct dephasing term

\begin{equation}
H_{z}=-\frac{i}{2}\varepsilon\left(t\right)\sin\left(\delta\phi\left(t\right)\right)\zeta_{a}\zeta_{s}.\label{eq:H_Z}
\end{equation}

As discussed in Section~\ref{sec:the-qubit-in}, static stray fields or static phase offsets do not cause decoherence; they simply shift the experimental tuning point required to reach the effective $\pi$-phase. However, dynamic fluctuations or jitter $\delta\phi(t)$ away from this ideal tuned condition introduce direct dephasing. 
The vulnerability of the qubit to this noise depends entirely on the
junction geometry and the physical mechanism used to create the $\pi$-phase.
It is important to stress the difference between \emph{long} and \emph{short}
junctions in this discussion. 

A \emph{long junction}, where the length of the normal region is much
larger than the superconducting coherence length, the two Majoranas
at the junction are bound to the superconductor-normal interfaces.
As shown in ~\citep[eq. (6)]{alicea2011nonabelian}, the hybridization
$\varepsilon\left(t\right)$ between these Majoranas is exponentially
suppressed by their separation. Hence, although the fluctuations on
$\delta\phi\left(t\right)$ exist, the drift in the energy of the
qubit is exponentially small in the case of a large junction. 

A \emph{short junction}, where the length of the normal region is
much smaller than the superconducting coherence length, does not have
the same spatial protection. Hence, in the absence of other protections,
the qubit should experience a large decoherence due to Eq.~(\ref{eq:H_Z}).

In the previous section I discussed three different physical implementations
for a $\pi$-junction:
\begin{itemize}
\item The first implementation relies mainly on the phase difference between
the two $s$-wave superconductors across the junction. 
\item The second implementation relies on a magnetic domain wall at the
junction. In this case the main source of dephasing from $\pi$ is
the change on magnetic environment. 
\end{itemize}
Both these proposals rely on a magnetic field inversion in the scale
of the junction, hence they are likely to only be realizable as \emph{long}
junctions. Hence, although $\phi\left(t\right)$ does exist, $\varepsilon\left(t\right)$
can be regarded as exponentially small.
\begin{itemize}
\item The third, and most physically feasible, implementation is  the direct
tunneling junction between wires with opposite spin-orbit orientations. 
\end{itemize}
Clearly this implementation corresponds to a \emph{short} junction.
The $\pi$ phase due to the relative orientations between the vectors 
$\hat{e}$ of both wires is always fixed. The only other noise source
would be from fluctuation in the magnetic field that produces the
Zeeman splitting. However, if the fluctuations are common to both
wires, they always preserve the relative $\pi$ phase at the junction.
Hence, neglecting high order effects due to gradients in the magnetic
field, this physical implementations of a $\pi$-junction is protected
from phase fluctuation by an \emph{inherent symmetry}. 

In summary, the decoherence term due to phase fluctuation, Eq.~(\ref{eq:H_Z}),
is exponentially small in the first two proposals and is \emph{absent}
in the third due to symmetry protection. Hence, in the following I
will not consider this term in qubit decoherence model.

\subsection{the microscopic noise model}

In the previous subsections, I established that the $\pi$-junction
qubit is vulnerable to quasiparticle poisoning from a \textquotedbl soft
gap''. I also argued that because the qubit's two Majorana partners,
$\zeta_{s}$ and $\zeta_{a}$, have distinct spatial profiles, they
must couple to independent, spatially-sorted sets of these Dynes states. 

By further including the effect of $1/f$ charge noise (which modulates
the coupling parameters via the wavefunction profiles, as discussed
above), the final microscopic noise Hamiltonian is:

\begin{align}
H_{\text{QP}} & =\sum_{i\in\text{Env}_{1}}\varepsilon_{i}f_{i}^{\dagger}f_{i}+\lambda_{i}\left(t\right)\left(f_{i}^{\dagger}-f_{i}\right)\zeta_{s}\nonumber \\
 & +\sum_{j\in\text{Env}_{2}}\varepsilon_{j}f_{j}^{\dagger}f_{j}+i\lambda_{j}\left(t\right)\left(f_{j}^{\dagger}+f_{j}\right)\zeta_{a}\label{eq:H_QP-1}
\end{align}

This Hamiltonian does not commute with the qubit's parity operator,
$P=-i\zeta_{s}\zeta_{a}$. The coupling to this fermionic environment
will, in general, cause the quantum information to decohere. To analyze
the effect of this environment, I must first understand its nature,
which depends on the number of environmental states $f_{i}$ that
significantly couple to the qubit.

The two Majoranas, $\zeta_{s}$ and $\zeta_{a}$, have spatially localized
wavefunctions. This localization means that I can explore two distinct
physical limits for their interaction with the continuum of Dynes
states.

\subsubsection{Discrete set of local levels}

It is conceivable that the Majorana wavefunctions are so sharply localized
that each one has a significant spatial overlap with only a handful
(or even one) of the environmental Dynes states. In this limit, the
noise spectrum is not smooth. Instead, the qubit is poisoned by Random
Telegraph Noise (RTN), as the environment is dominated by the random
\textquotedbl blinking\textquotedbl{} of a few specific defects {[}~\citep{kirton1989noise,simoen2016random}.
As I argued for the discrete ABS, such a situation might be manageable
if the chemical potential can be tuned to move that specific defect
out of resonance, but the outcome would be highly dependent on the
random, specific properties of the device.

\subsubsection{Dense set of local levels\protect \\
}

The opposite limit is to assume the Majoranas couple to a large number
of the environmental states. This is a very likely scenario, as the
\textquotedbl Dynes continuum\textquotedbl{} is formed from a vast
sea of disorder-induced states, both in the wire and in the underlying
$s$-wave superconductor. This dense set scenario presents a severe
problem, since, unlike a few discrete RTN sources, this collective
bath cannot be removed by gating. 

In the \emph{statistical} limit where the qubit interacts with the
entire ``Dynes continuum'', the noise from all the independent RTN
sources averages out. The resulting noise spectrum is no longer spiky
but becomes a smooth continuum. If the density of states at zero energy
is finite, then the system is well-described by the Dutta-Horn model
as $1/f$ noise ~\citep{dutta1981lowfrequency}. 

As discussed in Ref.~\citep{paladino20141}, the RTN in this continuum limit is
well-described within a Gaussian approximation, assuming a weak coupling between
the local level and the Dynes continuum. Under these conditions, the Central Limit
Theorem ensures that the noise experienced by the Majorana modes follows Gaussian
statistics. Since a Gaussian process is entirely characterized by its second-order
correlation function, the interaction is uniquely defined by the noise power
spectrum $S(\omega)$. For a zero-dimensional impurity problem, such as the Majorana modes considered here,
the specific statistics of the bath degrees of freedom
(fermionic vs. bosonic) are secondary to their collective spectral influence (see appendix~\ref{sec:fermion-boson-mapping} for a brief presentation).
Therefore, a bosonic bath is mathematically equivalent to a fermionic one,
provided it is constructed to reproduce the identical power spectrum $S(\omega)$.
While this equivalence can be formally derived using the path integral formalism
to integrate out the environmental degrees of freedom, the resulting effective
action depends only on the spectral density. This mapping from a fermionic
environment to a bosonic one is standard in quantum impurity problems, with
notable examples including the original discussions of decoherence in SQUIDs~\citep{leggett1984quantum}, superconducting
qubits~\citep{burkard2005circuit} and the Kondo
problem~\citep{affleck1991critical}.

There is a final key to modeling this type of bath as a bosonic environment,
which is a crucial property of our Hamiltonian in Eq. (\ref{eq:H_QP-1}).
Any perturbative expansion (such as a self-energy calculation) for these
environments involves evaluating correlation functions of the bath operators,
meaning the operators $f_{i}$ will always appear in pairs
(e.g., $\langle f_{i}(\tau)f_{i}^{\dagger}(0)\rangle$).
Because the bath fermions are non-interacting, it is straightforward
to show that the expansions are free of the fermionic sign problem.
This lack of interaction is precisely what allows the microscopic
fermionic environment to be mapped onto an effective bosonic one,
as the two systems belong to the same universality class.

Following the Caldeira-Leggett formalism~\citep{caldeira1983quantum,leggett1987dynamics},
any complex bosonic environment can be modeled in linear response
as an effective bath of harmonic oscillators. The noise power spectrum,
$S(\omega)$, is the main experimental measure that characterizes
the qubit environment. Via the fluctuation-dissipation theorem, this
is related to the spectral density, $J\left(\omega\right)$, by 

\begin{align}
S\left(\omega\right) & \propto\frac{J\left(\omega\right)}{\omega}\propto\omega^{s-1}.
\end{align}
In mesoscopic devices, like a Josephson junctions, the measured spectrum
is indeed the $1/f$ noise, corresponding to the $s\to0$ limit. This
\textquotedbl divergent\textquotedbl{} spectrum, $S(\omega)\propto1/\omega$,
is an extreme case of a sub-Ohmic environment.

This leaves a crucial ambiguity. The fact that the qubit's spatial
profile is exponentially localized makes it unclear what spectral
density $J(\omega)$ it truly couples. It is possible that the qubit
\emph{sees} the globally-averaged \emph{$1/f$} spectrum ($s\to0$)
of the entire device, or it may also be that the local nature results
in a different, perhaps less damaging, spectrum (e.g., an $s>0$).
\emph{Only an experimental measurement can settle this question and
therefore I will leave $s$ as an open parameter. }

By using the Pauli matrix representation for the Majoranas $\zeta_{a}\to\sigma^{z}$
and $\zeta_{s}\to\sigma^{x}$. It is possible to replace the original
Dynes environment by an effective bosonic bath. The resulting decoherence
model for the qubit is a small generalization of the usual spin-boson
model~\citep{leggett1987dynamics},

\begin{align}
H_{\text{eff}} & =\sum_{\omega}\omega a_{\omega}^{\dagger}a_{\omega}+i\frac{\lambda}{2}\sum_{\omega}{\cal F}\left(\omega\right)\left[a_{\omega}-a_{\omega}^{\dagger}\right]\sigma^{z}\nonumber \\
 & +\sum_{\omega}\omega b_{\omega}^{\dagger}b_{\omega}+i\frac{\lambda}{2}\sum_{\omega}{\cal F}\left(\omega\right)\left[b_{\omega}-b_{\omega}^{\dagger}\right]\sigma^{x},\label{eq:Q_frustration}
\end{align}
where I set $\hbar=1$, $\lambda$ is the dimensionless coupling constants,
\begin{align}
{\cal F}\left(\omega\right) & =\sqrt{\omega_{c}}\Omega_{c}^{\frac{1-s}{2}}\omega^{\frac{s}{2}}e^{-\frac{\omega}{2\Omega_{c}}},
\end{align}
$\omega_{c}$ and $\Omega_{c}$ are the infrared and ultra-violet
frequency cut-offs. In principle the two constants may be different,
however the averaging of the RTN couplings is likely to make them
the same. Therefore, I am assuming an emergent $U\left(1\right)$
symmetry in the noise model. Even if the couplings are not strictly
identical, the results from the $U\left(1\right)$ symmetric case persist until 
the energy scale of the asymmetry is reached."

The decoherence phenomenology of a single two level system couple
to a bosonic environment is well known. 
\begin{enumerate}
\item For $s>1$ the environment is known as super-Ohmic~\citep{lopez-delgado2017longtime}.
In this regime the environment leads to underdamping the two level system.
There are always coherent oscillations and the dynamics is perturbative
in the dimensionless coupling constant between the two level system
and the environment. All these facts can be synthesized in the renormalization
group equations, that in this case have an irrelevant flow. 
\item The case $s=1$ is known as the Ohmic environment~\citep{novais2005frustration}.
The renormalization equation is marginal and the fate of the two
level system will depend on value of the dimensionless coupling constant. 
\item Finally, for $0<s<1$ the environment is known as sub-Ohmic. A two-level
system coupled to a single sub-Ohmic bath is a famously \emph{doomed}
system. The bath's overwhelming density of low-frequency modes causes
total decoherence at any finite temperature~\citep{leggett1987dynamics}.
The system is always in the overdamped, incoherent regime. In the
renormalization group language the flow is to strong coupling and
it is not possible to describe perturbatively in the dimensionless
coupling constant the dynamics of the two level system.
\end{enumerate}
The situation is dramatically different for two environments, as I
will present in the next section.

\section{The qubit decoherence \label{sec:decoherence}}

The noise sources discussed in the previous section present two very
distinct possibilities to the decoherence in a $\pi$-junction qubit.
If there is only a few environmental states interacting with the qubit,
then the decoherence would be manageable by gating the levels out
of resonance. Conversely, if there is a large set of environmental
states interacting with the qubit, the effective noise model is a
generalized spin-boson model.

To fully understand the decoherence in the spin-boson model it is
instructive to set aside the full Hamiltonian, Eq.~(\ref{eq:Q_frustration}),
and first analyze one of its components: the pure dephasing model~\citep{unruh1995maintaining,breuer2010theory}.

\subsection{the pure dephasing model}

This simplified model is exactly solvable and provides a clear picture
of decoherence. It describes the qubit coupling to only one of the
baths, with no spin-flip terms~\citep{unruh1995maintaining,breuer2010theory},

\begin{equation}
H_{d}=\sum_{\omega}\omega a_{\omega}^{\dagger}a_{\omega}+i\frac{\lambda}{2}{\cal F}\left(\omega\right)\left[a_{\omega}-a_{\omega}^{\dagger}\right]\sigma^{z},\label{eq:pure-dephasing}
\end{equation}
where ${\cal F}\left(\omega\right)=\sqrt{\omega_{c}}\Omega_{c}^{\frac{1-s}{2}}\omega^{\frac{s}{2}}e^{-\frac{\omega}{2\Omega_{c}}}$.

This model can be diagonalized using a generalized polaronic rotation,
\begin{equation}
P=e^{-i\frac{\lambda}{2}\sum_{\omega}{\cal G}\left(\omega\right)\left[a_{\omega}+a_{\omega}^{\dagger}\right]\sigma^{z}},
\end{equation}
where ${\cal G}\left(\omega\right)=\sqrt{\omega_{c}}\Omega_{c}^{\frac{1-s}{2}}\omega^{\frac{s}{2}-1}e^{-\frac{\omega}{2\Omega_{c}}}$.
This transformation exactly cancels the interaction term, leaving
a free Hamiltonian,

\begin{align}
\bar{H}_{d} & =PH_{d}P^{\dagger},\nonumber \\
 & =\sum_{\omega}\omega a_{\omega}^{\dagger}a_{\omega}-\frac{\lambda^{2}}{4}\omega_{c}\sum_{\omega}\Omega_{c}^{1-s}\omega^{s-1}e^{-\frac{\omega}{\omega_{c}}}.\label{eq:rotatedH_d}
\end{align}
The exact evolution operator is simply

\begin{equation}
U\left(t\right)=P^{\dagger}e^{-i\sum_{\omega}\omega ta_{\omega}^{\dagger}a_{\omega}}P.
\end{equation}
If I prepare an initial superposition state, $\left|\psi_{0}\right\rangle =\left(a\left|\uparrow\right\rangle +b\left|\downarrow\right\rangle \right)\otimes\left|0\right\rangle $,
the off-diagonal terms of the qubit's reduced density matrix decay
as

\begin{equation}
\rho_{\uparrow\downarrow}\left(t\right)=a b^{*}I\left(t\right),
\end{equation}
The decoherence function $I\left(t\right)$ contains all the non-trivial
dynamics. It is given by

\begin{equation}
I\left(t\right)=e^{-\lambda^{2}\sum_{\omega}\omega_{c}\Omega_{c}^{1-s}\omega^{s-2}e^{-\frac{\omega}{\omega_{c}}}\left[1-e^{-i\omega t}\right]},
\end{equation}
and in the continuum limit is a well known integral, which its properties
depend on the value of $s$:
\begin{itemize}
\item super-Ohmic case, $s-1>1$, the integral converges and leads to
\begin{equation}
I\left(t\right)=\exp\left[-\lambda^{2}\Gamma\left(s-1\right)\left(\frac{1}{2^{s-1}}\right)\left[1-\left(1+i\frac{\Omega_{c}t}{2}\right)^{-\left(s-1\right)}\right]\right].\label{eq:decoherence-superohmic}
\end{equation}
Hence, in the long time limit the coherence of the reduced density
matrix are finite.
\item Ohmic case, $s=1$, the coherences, $\rho_{\uparrow\downarrow}\left(t\right)$,
go to zero in the long time limit as a power law,
\begin{equation}
I\left(t\right)=\frac{1}{\left|1+i\frac{\Omega_{c}t}{2}\right|^{\lambda^{2}}}.\label{eq:decoherencce-ohmic}
\end{equation}
\item sub-Ohmic case, $0<s<1$, the integral can be obtained by analytical
continuation and gives a result that formally is identical to the
super-Ohmic case, Eq.~(\ref{eq:decoherence-superohmic}), but the
coherence is now exponentially suppressed in the long time limit.
\item In the $1/f$ limit, $s\to0$, the $\Gamma$ function diverges, $\lim_{s\to0^{+}}\Gamma\left(s-1\right)\approx-\frac{1}{s}$.
The system decoheres almost instantaneously.
\end{itemize}
The exact solution shows what happens, but a different approach, mapping
to a classical model, shows the underlying reason why. 

The simplest possible calculation is to evaluate the partition function
of Eq.~(\ref{eq:pure-dephasing}), which is equivalent to the evolution
operator by a Wick rotation. Using a Trotter-Suzuki path-integral
expansion in imaginary time, one can integrate out the bosonic fields.
This maps the quantum dephasing model, Eq.~(\ref{eq:pure-dephasing}),
onto a 1D classical Ising model with long-range, ferromagnetic interactions~\citep{anderson1969exact}

\begin{equation}
\frac{Z_{d}\left(\lambda\right)}{Z_{0}}=\sum_{\left\{ \sigma\right\} }e^{-\lambda^{2}\int_{-\infty}^{\infty}d\tau\int_{-\infty}^{\tau_{1}}d\tau_{2}G\left(\tau_{1},\tau_{2}\right)\sigma^{z}\left(\tau_{1}\right)\sigma^{z}\left(\tau_{2}\right)},
\end{equation}
where $Z_{0}$ is the free boson partition function, $\sigma^{z}\left(\tau\right)=\pm1$
is a classical variable that represents the qubit's state at imaginary
time $\tau$, and
\begin{align}
G\left(\tau_{1},\tau_{2}\right) & =\ensuremath{\frac{\Gamma(s+1)}{2^{s+1}}\Omega_{c}^{2}\left(1+\frac{\Omega_{c}(\tau_{1}-\tau_{2})}{2}\right)^{-(s+1)},}\nonumber \\
 & \approx\ensuremath{\Gamma(s+1)\omega_{c}^{2}\left[\omega_{c}(\tau_{1}-\tau_{2})\right]^{-(s+1)}.}\label{eq:propagator}
\end{align}
For this particular Ising model there is no entropic cost in producing
a domain wall. In the limit that $\lambda\to0$ all spin configurations
have exactly the same statistical weight and the system is in a paramagnetic
phase. It is a well known result that this model has a transition
to a ferromagnetic ordered phase for interactions that decays slower
than $1/\left(\tau_{1}-\tau_{2}\right)^{2}$. In other words, for
$s>1$ the system is in a paramagnetic phase where domain walls are
deconfined, whereas for $s<1$ the system is in a ferromagnetic phase
and the domain walls are confined. 

This thermodynamic phase diagram provides a beautiful insight on why
the sub-Ohmic bath is so detrimental to quantum information, since
the evolution operator is mathematically equivalent to a system that
always orders itself, freezing out the quantum dynamics. This sets
the stage for our central question: what happens when I add the second,
non-commuting bath from Eq.~(\ref{eq:Q_frustration})?

\subsection{Quantum frustration of decoherence}

The pure dephasing model is a text book example of decoherence. It
shows that an environment can destroy a superposition, thus defining
a ``pointer basis''~\citep{zurek2000einselection,zurek2003decoherence}.
In the case discussed in the previous paragraph, the pointer basis
was the $\sigma^{z}$ eigenvectors. This mechanism is one of the most
accepted paths to the emergence of classicality in a quantum world. 

From this perspective, the long-time behavior of our full model, Eq.~(\ref{eq:Q_frustration}),
is not obvious. The two environments try to define two incompatible
pointer basis, $\sigma^{z}$ and $\sigma^{x}$, simultaneously. The
baths frustrate each other. This mechanism is called quantum frustration
of decoherence~\citep{castroneto2003quantum,novais2005frustration,feller2020einselection}.

Quantum frustration was initially studied in the context of magnetic
impurities in magnetic environments~\citep{castroneto2003quantum,novais2005frustration}.
In that original case the two environments were Ohmic, $s=1$, and
the natural language to describe the phase transition was the renormalization
group (RG). The final result for the Ohmic case of Eq.~(\ref{eq:Q_frustration})
is the RG equations

\begin{align}
\frac{\partial\lambda}{\partial\ell} & =-\lambda^{3},
\end{align}
where $d\ell=-d\Omega/\Omega$ is the renormalization flow parameter. 

The most direct path to derive this RG equation is to consider the
same polaronic rotation that I used in the pure dephasing model,

\begin{align}
\bar{H}_{\text{eff}} & =PH_{\text{eff}}P^{\dagger}\nonumber \\
 & =\sum_{\omega}\omega a_{\omega}^{\dagger}a_{\omega}+\sum_{\omega}\omega b_{\omega}^{\dagger}b_{\omega}\nonumber \\
 & +i\frac{\lambda}{2}\sum_{\omega}{\cal F}\left(\omega\right)\left[b_{\omega}-b_{\omega}^{\dagger}\right]e^{-i\lambda\sum_{\omega}{\cal G}\left(\omega\right)\left[a_{\omega}+a_{\omega}^{\dagger}\right]}\sigma^{+}\nonumber \\
 & +\text{h.c.}-\frac{\lambda^{2}}{4}\omega_{c}\sum_{\omega}\Omega_{c}^{1-s}\omega^{s-1}e^{-\frac{\omega}{\omega_{c}}}.\label{eq:rotated_H_eff}
\end{align}

The partition function can once again be mapped into a 1D Ising model
with long range interactions, where the interaction between \emph{domain
walls} in the Ising model decays as a power law of their time difference,
$\sim\lambda^{2}/\left(\tau_{1}-\tau_{2}\right)^{2\left(1+\lambda^{2}\right)}.$
This is a similar behavior to the super-Ohmic environment in the pure
dephasing model, $\sim\lambda^{2}/\left(\tau_{1}-\tau_{2}\right)^{s-1}$. 

The ground state degeneracy is $\ln2$ indicating that the entanglement
between the qubit and the environment is always small. In terms of
the dynamics, the long time behavior of the model no longer has the
coherence going to zero~\citep{castroneto2003quantum,novais2005frustration}. 

The $s<1$ case is highly nontrivial. The domain wall interactions
are now exponentially decaying,

\begin{equation}
\sim\lambda^{2}\frac{e^{-\lambda^{2}\left(\tau_{1}-\tau_{2}\right)^{\left(1-s\right)}}}{\left(\tau_{1}-\tau_{2}\right)^{2}}n,
\end{equation}
but the dissipation strength $\lambda$ also defines a scale in the
problem that hinders the simple analyses that I did for the previous
cases (ferromagnetic or paramagnetic phase). 

It is possible to extend the renormalization group analyses using
an $\varepsilon$-expansion formalism to other values of $s$. By
considering $1-s$ an infinitesimal parameter, it is straightforward
to derive the equation RG equation~\citep{guo2012critical,bruognolo2014twobath}

\begin{equation}
\frac{\partial\lambda}{\partial\ell}=\left(1-s\right)\lambda-\lambda^{3}.
\end{equation}
For the sub-Ohmic case, $s<1$, this equation indicates that there
is an intermediate fixed point at $\lambda_{c}=\sqrt{1-s}$, but since
this value is finite, the pertubative calculation is not trustworthy.

For a long time it was assumed that this intermediate phase would
exists for all values of $s<1$. However, after more careful numerical
studies~\citep{guo2012critical,vojta2012numerical,bruognolo2014twobath}
it was found a much more complex situation.

Significantly, numerical studies show that for $s<s^{*}\sim0.76\pm0.01$ the model
goes to a \emph{localized} phase. This phase corresponds to an spontaneous
symmetry breaking of the $U\left(1\right)$ symmetry of the noise
model. The expectation value of $\sigma^{x}$ and $\sigma^{z}$ are
both not zero, indicating that the quantum fluctuations due to the
noncommutative nature of the couplings was not enough to keep the
system from choosing a pointer basis. All qubit operators in the $\left\{ x,z\right\} $
plane connected by the $U\left(1\right)$ symmetry of the noise model
will have the same expectation value. Hence, for $s<s^{*}$ the quantum
information in the qubit will be lost.

For $s>s^{*}$ and small $\lambda$, there is an intermediate phase
that corresponds to the intermediate fixed point found in the perturbative
RG. It is known that this fixed point has a non-zero ground state
entropy smaller than $\ln2$. This means that the qubit in the long
time regime is partially entangled with the environment, but there
is no catastrophic lost of coherence~\citep{guo2012critical,vojta2012numerical,bruognolo2014twobath}.

\section{Dicussion and conclusions\label{sec:Dicussion-and-conclusions}}

In this manuscript, I presented the possible physical realizations
of a $\pi$-junction qubit, which is encoded by two co-located Majorana
modes. Although this is explicitly not a topologically protected qubit,
it benefits from a similar non-local principle: the two Majoranas
have distinct spatial profiles, meaning each one is coupled to an
independent environmental bath.

The main challenge is quasiparticle poisoning (QP) from disorder-induced
local levels. The viability of the qubit depends on the effective
number of these levels coupled to the Majoranas.

On one hand, if the qubit interacts with only a small
number of discrete local levels, the decoherence can be managed by
gating. On the other hand, if the number of coupled levels is large,
a statistical average occurs. Gating fails because as I tune some
states out of resonance, others are simultaneously brought into it.
This scenario maps directly to the dangerous sub-Ohmic spin-boson
model.

In this catastrophic limit, the effective noise model is a generalized
spin-boson model, characterized by the noise power spectrum $S(\omega)\propto\omega^{s-1}$.
The parameter $s$ determines the qubit's fate via the mechanism of
quantum frustration of decoherence:
\begin{itemize}
\item $s>1$ (super-Ohmic); the qubit dynamics is perturbative in the dimensionless
coupling constant with the environment. The qubit entropy is $\ln2$,
as it remains disentangled from the environment.
\item $s=1$ (Ohmic); the quantum frustration of the two environments keeps
the qubit in the perturbative regime, with its entropy still $\ln2$.
\item $0.76<s<1$ (sub-Ohmic); quantum frustration partially protects the
qubit. In the long time limit the qubit partially entangles with the
environment and its entropy is $<\ln2$.
\item $0\leq s<0.76$ (sub-Ohmic); there is a spontaneous symmetry breaking
of the $U\left(1\right)$ noise model. In the long time limit the
qubit entangles with the environments and its entropy is zero. 
\end{itemize}
Determining $s$ that describe the environment of a large set of local
levels is an open experimental question. If the density of such states
at zero energy is constant, then $s\to0^{+}$ and the qubit would
always suffer catastrophic decoherence. However, the local spatial
profile of the qubit limits the number of in-gap states that interact
with it. It is conceivable that the number of] local states interacting
with the qubit at zero energy to be zero, or made zero by gating. 

If the local coupling to zero-energy states is eliminated, the qubit's
coherence would be inherently protected by the quantum frustration
mechanism, making this a viable path for solid-state qubits.
\begin{acknowledgments}
This work was partially supported by the S\~{a}o Paulo Research Foundation
(FAPESP), Brazil, Process No. 2022/15453-0. The author would like
to thank Luis G. G. V. Dias da Silva for insightful discussions.
\end{acknowledgments}

\appendix

\section{\label{sec:diagonalizing-the-free}Diagonalizing the free fermion
problem}

In this appendix, I sketch the analytical solution for the free fermion
problem following the Lieb-Schulz-Mattis method~\citep{lieb1961two}.
Using the Nambu notation
\begin{equation}
\vec{v}^{\dagger}=\left[\begin{array}{cccccc}
\eta_{-\frac{N}{2}} & \cdots & \eta_{\frac{N}{2}-1} & \nu_{-\frac{N}{2}} & \cdots & \nu_{\frac{N}{2}-1}\end{array}\right],
\end{equation}
the Hamiltonian Eq.~(\ref{eq:majorana-model}) can be written as

\begin{equation}
H_{\pi}=\frac{1}{4}\vec{v}^{\dagger}\left[\begin{array}{cc}
0 & iM\\
-iM^{t} & 0
\end{array}\right]\vec{v}
\end{equation}
with

\begin{equation}
M=\left[\begin{array}{ccccc}
\mu & \beta_{-\frac{N}{2}} & 0 & \cdots & 0\\
\alpha_{-\frac{N}{2}} & \mu & \beta_{-\frac{N}{2}+1} & \cdots & 0\\
0 & \alpha_{-\frac{N}{2}+1} & \mu &  & 0\\
\vdots & \vdots &  &  & \vdots\\
0 & 0 &  & \ddots & \beta_{\frac{N}{2}-2}\\
0 & 0 & \cdots & \alpha_{\frac{N}{2}-2} & \mu
\end{array}\right].
\end{equation}
The Schr\"odinger equation that needs to be solved is

\begin{equation}
\left[\begin{array}{cc}
0 & iM\\
-iM^{t} & 0
\end{array}\right]\left[\begin{array}{c}
\vec{\phi}_{k}\\
i\vec{\xi}_{k}
\end{array}\right]=\Lambda_{k}\left[\begin{array}{c}
\vec{\phi}_{k}\\
i\vec{\xi}_{k}
\end{array}\right],\label{eq:single_particle}
\end{equation}
where $\vec{\phi}$ and $\vec{\xi}$ are real vectors. 

If $\Lambda_{k}\neq0$ the problem reduces to solve the eigenvalue
problem 

\begin{equation}
MM^{t}\vec{\phi}_{k}=\Lambda_{k}^{2}\vec{\phi}_{k},
\end{equation}
that leads to the equation

\begin{widetext}

\begin{equation}
\alpha_{n}\beta_{n+1}\phi_{n+2}^{k}+\mu\left(\alpha_{n}+\beta_{n}\right)\phi_{n+1}^{k}+\left(\alpha_{n}^{2}+\beta_{n-1}^{2}+\mu^{2}\right)\phi_{n}^{k}+\mu\left(\alpha_{n-1}+\beta_{n-1}\right)\phi_{n-1}^{k}+\beta_{n-1}\alpha_{n-2}\phi_{n-2}^{k}=\Lambda_{k}^{2}\phi_{n}^{k},
\end{equation}
\end{widetext}That in general can be solved using the Ansatz
\begin{equation}
\phi_{n}^{k}=a_{n}\cos\left(kn\right)+b_{n}\sin\left(kn\right)
\end{equation}
and lead to the band solutions of the model. 

The states that are on the kernel of $M$ need to be solved separately,
and they correspond to states with $\Lambda_{k}=0$ that do not belong
to the bands. Considering the equations

\begin{equation}
M^{t}\vec{\phi}=0,
\end{equation}
correspond to solve the coupled equations, 

\begin{equation}
\alpha_{n-1}\phi_{n-1}+\mu\phi_{n}+\beta_{n}\phi_{n+1}=0.
\end{equation}

As a possible example of a short junction, consider
\begin{equation}
\left(\alpha_{n},\beta_{n}\right)=\begin{cases}
\left(1-\gamma,1+\gamma\right) & n\leq-2,\\
\left(t-\upsilon,t+\upsilon\right) & n=-1,\\
\left(t+\upsilon,t-\upsilon\right) & n=0,\\
\left(1+\gamma,1-\gamma\right) & n\geq1,
\end{cases}
\end{equation}
where $t<1$ is the tunneling amplitude between the segments of the
wire and $\upsilon<\gamma$ is a transition pairing strength. 

There are four zero energy modes in this model, two at the edges and
two in the junction.

For an infinite wire $N\to\infty$ and the Kitaev limit, $\gamma=\pm1$,
it is possible to write a simple analytical solution to the zero energy
states at junction region. There are two Majorana modes with zero
energy, a symmetric and antisymmetric mode 

\vspace{0.2in}

\begin{widetext}
\begin{align}
\zeta_{s} & =a_{1}\left\{ \eta_{0}-\frac{\mu}{2\left(t-\upsilon\right)}\left(\eta_{1}+\eta_{-1}\right)+\left(\frac{\left(t+\upsilon\right)}{2}-\frac{\mu^{2}}{4\left(t-\upsilon\right)}\right)\left(\eta_{2}+\eta_{-2}\right)+\sum_{n=3}^{\infty}\frac{\left(-\mu\right)^{n-2}}{2}\left(\frac{\left(t+\upsilon\right)}{2}-\frac{\mu^{2}}{4\left(t-\upsilon\right)}\right)\left(\eta_{n}+\eta_{-n}\right)\right\} ,\\
\zeta_{a} & =a_{2}\sum_{n=1}^{\infty}\frac{\left(-\mu\right)^{n-1}}{2}\left(\eta_{n}-\eta_{-n}\right)
\end{align}
\end{widetext}where $a_{1,2}$ are normalization constants. This
solution clearly shows the two very different spatial profiles that
each Majorana has.

\section{\label{sec:Semiconductor-quantum-wires}Semiconductor quantum wires
and the Kitaev chain}

For completeness, in this appendix I follow the discussion from Refs.~\citep{lutchyn2010majorana,oreg2010helical,klinovaja2012transition,alicea2012new}
on how to map a physical system to the idealized Kitaev chain model.

The minimal model that would capture the physics is to consider a
semiconductor wires with band structure

\begin{equation}
H_{b}=\int dx\vec{\psi}^{\dagger}\left(x\right)\left[-\frac{\hbar^{2}\partial_{x}^{2}}{2m}-\mu-i\alpha\vec{e}.\vec{\sigma}\tau^{z}\partial_{x}\right]\vec{\psi}\left(x\right),
\end{equation}
where $m$ is the effective electron mass, $\alpha$ is the spin-orbit
coupling, $\hat{e}$ is the spin-orbit orientation with respect to
the underlying crystalline structure, $\left\{ \vec{\sigma},\vec{\tau}\right\} $
are the Pauli matrices acting on the spin and electron-hole space
respectively, $\mu$ is a chemical potential that can be controlled
by an external electrostatic gate, and 
\begin{equation}
\vec{\psi}^{\dagger}\left(x\right)=\left[\begin{array}{cccc}
\psi_{\uparrow}^{\dagger}\left(k\right) & \psi_{\downarrow}^{\dagger}\left(k\right) & \psi_{\uparrow}\left(k\right) & \psi_{\downarrow}\left(k\right)\end{array}\right]
\end{equation}
are the standard fermionic operators in the Nambu notation. It is
usually assumed that $\hat{e}=\hat{y}$~\citep{alicea2012new}, hence
the wire Hamiltonian is 

\begin{equation}
H_{b}=\int dx\vec{\psi}^{\dagger}\left(x\right)\left[-\frac{\hbar^{2}\partial_{x}^{2}}{2m}-\mu-i\alpha\sigma^{y}\tau^{z}\partial_{x}\right]\vec{\psi}\left(x\right).
\end{equation}

\begin{figure}
\includegraphics[width=0.75\columnwidth]{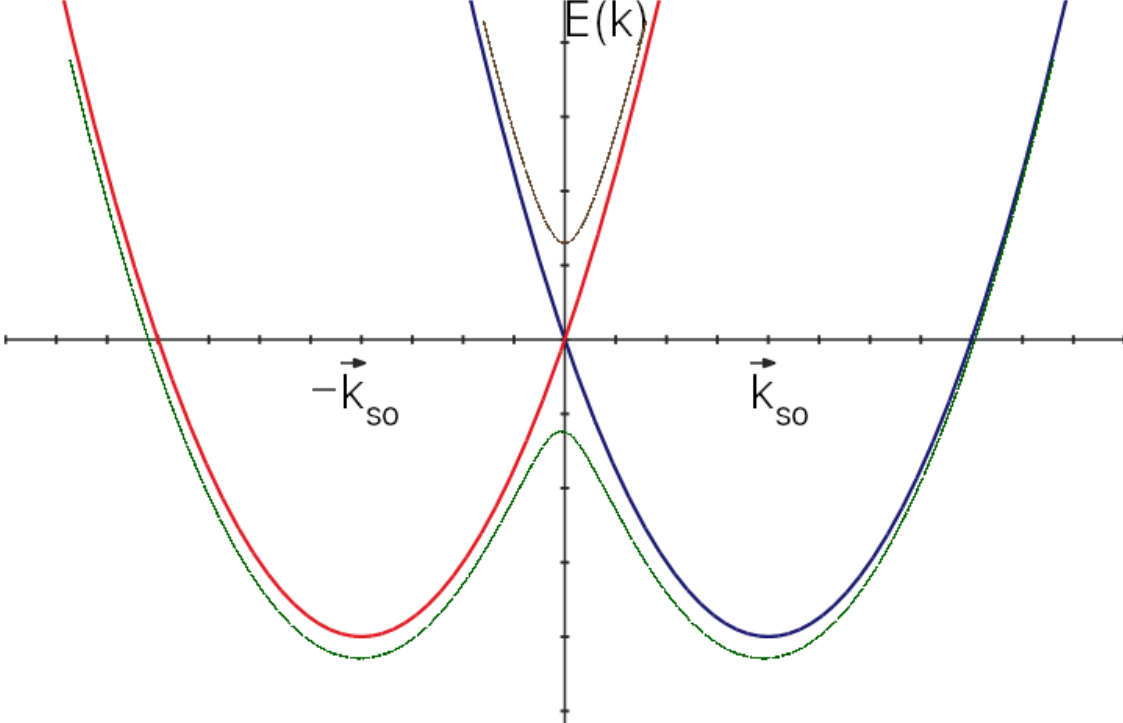}

\caption{\label{fig:A-simple-1D-band}A simple 1D wire bands with Rashba spin
orbit interaction. The blue/red lines correspond to the two spin bands
that are dislocated by the spin-orbit vector $\left|\vec{k}_{so}\right|=m\alpha/\hbar^{2}$.
After introducing the superconducting pairing the two band hybridize
creating the green and brown bands. At $k=0$ the gap between the
hybridize bands is $\left|\Delta\right|$.}
\end{figure}

The spin-orbit coupling moves the spin up/down bands in momentum space
by the vector $k_{so}=m\alpha/\hbar^{2}$, see Fig. \ref{fig:A-simple-1D-band},
and for simplicity the chemical potential is chosen to be zero at
the points where the Rashba branches touch at the $k=0$ momentum.

These nanowires are deposited on top of an $s$-wave superconductor,
commonly $Nb$ or $Al$, so that the superconducting condensate induces
a finite pairing amplitude 

\begin{equation}
H_{s}=\int dx\Delta\psi_{\uparrow}^{\dagger}\left(x\right)\psi_{\downarrow}^{\dagger}\left(x\right)+\text{h.c.},
\end{equation}
in the wire through the proximity effect~\citep{alicea2012new}. This
hybridizes the two parabolic bands, leading to the dispersion

\begin{equation}
E\left(k\right)=k^{2}+k_{so}^{2}\pm\sqrt{\left(2k_{SO}k\right)^{2}+\left|\Delta\right|^{2}},
\end{equation}
that has a band gap at $k=0$ proportional to $\left|\Delta\right|$. 

To complete the mapping onto the Kitaev chain, a magnetic field is
applied perpendicular to the spin-orbit vector $\hat{e}$. If I assume
again that $\hat{e}=\hat{y}$ a possible choice is
\begin{equation}
H_{z}=\int dx\vec{\psi}^{\dagger}\left(x\right)\left[-\frac{1}{2}g\mu_{B}B_{0}\sigma^{z}\tau^{z}\right]\vec{\psi}\left(x\right).\label{eq:H_z}
\end{equation}

When the Zeeman splitting dominates over both the induced superconducting
pairing and the chemical potential, the higher-energy spin sector
becomes frozen out, leaving an effective spinless model. In this low-energy
limit, the Hamiltonian of the nanowire coincides with Eq.~(\ref{eq:Kitaev})~\citep{alicea2012new}.

\section{\label{sec:fermion-boson-mapping}Mapping the Fermionic Environment to an Effective Bosonic Bath}

This appendix formally justifies the transition from the microscopic fermionic Hamiltonian, Eq.~(\ref{eq:H_QP-1}) , to the effective spin-boson model, Eq.~(\ref{eq:Q_frustration}). Here, I follow the traditional approach developed by Caldeira-Leggett\cite{caldeira1983quantum} using the path integral formalism. 

I begin by representing the two local Majorana modes using Pauli matrices, $\zeta_s \rightarrow \sigma^x$ and $\zeta_a \rightarrow \sigma^z$. Applying this substitution to the interaction terms in Eq.~(\ref{eq:H_QP-1}) yields a system-environment coupling of the form
\begin{equation}
H_{int} = \sigma^x X_1 + \sigma^z X_2,
\end{equation}
where the bath operators are defined as $X_1 = \sum_{i \in Env_1} \lambda_i(t)(f_i^\dagger - f_i)$ and $X_2 = i\sum_{j \in Env_2} \lambda_j(t)(f_j^\dagger + f_j)$.

In the path-integral formalism, the partition function of the qubit coupled to the environment is expressed as an integral over both the qubit paths $\sigma(\tau)$ and the environmental Grassmann variables $\bar{f}, f$

\begin{equation}
	Z = \int \mathcal{D}[\sigma] \mathcal{D}[\bar{f}, f] e^{-\frac{1}{\hbar}(S_{sys} + S_{bath} + S_{int})}
\end{equation}

I can write the interaction action in imaginary time $\tau$ as a linear coupling

\begin{equation}
S_{int} = \int_0^{\hbar \beta} d\tau \left[ \sigma^x(\tau) X_1(\tau) + \sigma^z(\tau) X_2(\tau) \right],
\end{equation}
where the time-dependent bath operators naturally follow from their Hamiltonian definitions.

Because the unperturbed environment consists of non-interacting fermions, the bath action $S_{bath}$ is strictly quadratic in the Grassmann variables and obeys Wick's theorem. This quadratic nature allows us to trace out the environmental degrees of freedom exactly via Gaussian integration. This procedure yields the reduced partition function for the qubit, characterized by the Feynman-Vernon influence functional $\mathcal{F}[\sigma]$,
\begin{equation}
Z_{red} = \int \mathcal{D}[\sigma] e^{-\frac{1}{\hbar} S_{sys}} \mathcal{F}[\sigma].	
\end{equation}

The exact evaluation of this Gaussian integral produces an influence functional that depends exclusively on the two-point correlation functions of the bath operators (a consequence of Wick's theorem for non-interacting particles)

\begin{widetext}

\begin{equation}
\mathcal{F}[\sigma] = \exp\left[ \frac{1}{2\hbar^2} \int_0^{\hbar \beta} d\tau \int_0^{\hbar \beta} d\tau' \sum_{\alpha \in \{x,z\}} \sigma^\alpha(\tau) \langle X_\alpha(\tau) X_\alpha(\tau') \rangle_0 \sigma^\alpha(\tau') \right],	
\end{equation}

\end{widetext}
where I specifically assumed that the two environments are independent, enforcing that the cross-correlation functions vanish, $\langle X_x(\tau) X_z(\tau') \rangle_0=0$.

The correlation function $\langle X_\alpha(\tau) X_\alpha(\tau') \rangle_0$ is the Fourier transform of the bath's noise power spectrum, $S(\omega)$. Because the influence functional is completely determined by this two-point correlator, the specific quantum statistics of the underlying bath particles, whether fermions or bosons, is mathematically irrelevant to the qubit's reduced dynamics. Consequently, I can replace the original fermionic continuum with an effective bath of non-interacting harmonic oscillators, Eq.~(\ref{eq:Q_frustration}), provided I tune the bosonic spectral density $J(\omega)$ to reproduce the exact same noise spectrum, $S(\omega)$, of the fermionic environment.

\section{\label{sec:independent-baths}Spatial Orthogonality and the Independence of Environmental Baths}

To formally justify the assumption that the $\pi$-junction qubit couples to two effectively independent environments, I must analyze the spatial profiles of the Majorana modes and their interaction with the dense continuum of local defects. The two zero-energy Majorana modes at the junction, $\zeta_s$ and $\zeta_a$, combine to form a single conventional fermionic state. Because these modes are distinct eigenstates of the underlying Bogoliubov-de Gennes (BdG) Hamiltonian, linear algebra dictates that their spatial wave functions must be strictly orthogonal.

How this required orthogonality manifests physically depends entirely on the geometry of the junction. In a long junction, where the separation between the superconducting segments is much larger than the coherence length ($L \gg \xi$), the Majorana modes are bound to the opposing Superconductor-Normal (SN) interfaces. Because their wave functions decay exponentially into the normal region, they are spatially separated. In this long-junction limit, their spatial overlap is exponentially suppressed, which trivially satisfies the orthogonality constraint. In this case, the independence of the bath can be readily argued by the spatial separation between the Majoranas (although the argument below can also be applied to this case). 

In a short junction ($L \ll \xi$), however, the Majorana modes are co-located at the center of the junction and cannot rely on spatial separation. To maintain  orthogonality while confined to the exact same spatial region, the two modes are forced to adopt highly distinct spatial profiles. As explicitly derived for the effective Kitaev chain in Appendix~\ref{sec:diagonalizing-the-free}, this results in one mode exhibiting a symmetric spatial distribution (peaking at the junction center), while the other must exhibit an antisymmetric distribution (possessing a node at the center).

To see how this spatial distinction decouples the environmental baths, it is instructive to decompose the microscopic bath fermions into their Majorana components, $f_k = \frac{1}{2}(\eta_k + i\nu_k)$. Substituting this into the interaction terms of Eq.~(\ref{eq:H_QP-1}), the coupling to $\zeta_s$, proportional to $(f_k^\dagger - f_k)$, reduces strictly to an interaction with the $\nu_k$ bath Majorana. Conversely, the coupling to $\zeta_a$, proportional to $i(f_k^\dagger + f_k)$, reduces strictly to an interaction with the $\eta_k$ bath Majorana. Algebraically, the two local modes initially couple to completely independent degrees of freedom of the environmental fermions.

However, the free Hamiltonian of the local levels, $H_{bath} = \sum_k \epsilon_k f_k^\dagger f_k$, couples their evolution
\begin{equation}
H_{bath} =\sum_k \frac{\epsilon_k}{2} (1 + i \eta_k \nu_k),
\end{equation}
meaning that the $\eta_k$ and $\nu_k$ modes have a non-zero correlation over time,
\begin{equation}
\langle \nu_k(t) \eta_k(0) \rangle \neq 0.
\end{equation} 
In any perturbative expansion of the qubit's dynamics, this non-zero correlator would typically induce correlated noise between $\zeta_s$ and $\zeta_a$.

The key to realizing that the qubit couples to independent environments lies in the fact that the Majoranas have orthogonal wave functions and the continuous nature of the Dynes environment.

The total cross-correlation between the two environmental Majoranas always has the form
\begin{equation}
\sum_k \lambda_{s,k} \lambda_{a,k} \langle \nu_k(t) \eta_k(0) \rangle,
\end{equation}
where I explicitly used that the environmental fermions obey Wick's theorem. Crucially, the expectation value $\langle \nu_k(t) \eta_k(0) \rangle$ depends on the energy $\epsilon_k$ of the mode, meaning it cannot be factored out of the sum.

To evaluate this sum over the dense defect continuum, I must express the microscopic coupling constants in terms of their spatial overlaps. Let $\phi_k(x)$ represent the localized spatial wave function of the environmental defect state associated with the fermion operator $f_k$. The coupling magnitude $\lambda_{\alpha,k}$ (where $\alpha \in \{s, a\}$) is dictated by the spatial overlap between the local Majorana wave function $\zeta_\alpha(x)$ and this bath profile 
\begin{equation}
\lambda_{\alpha,k} \propto \int dx \, \zeta_\alpha(x) \phi_k(x)
\end{equation}
The full macroscopic cross-correlation therefore involves a sum over $k$ of a double spatial integral mixed with the energy-dependent dynamic term. Taking the continuum limit, I replace the sum over discrete modes $k$ with a joint integral over energy $\epsilon$ and space $x$, weighted by the energy-dependent local density of states of the bath, $\rho(x, \epsilon)$\begin{widetext}
\begin{equation}
\sum_k \lambda_{s,k} \lambda_{a,k} \langle \nu_k(t) \eta_k(0) \rangle \propto \int d\epsilon \int dx \int dx' \, \zeta_s(x) \zeta_a(x') \rho(x, \epsilon) \delta(x - x') \langle \nu(\epsilon, t) \eta(\epsilon, 0) \rangle,
\end{equation}\end{widetext}
where the spatial delta function arises from the random phases of the dense, non-interacting, defect ensemble, eliminating non-local interference terms ($x \neq x'$).
Hence, the cross-correlation becomes
\begin{equation}
\int d\epsilon \int dx \, \zeta_s(x) \zeta_a(x) \rho(x, \epsilon) \langle \nu(\epsilon, t) \eta(\epsilon, 0) \rangle.
\end{equation}

To proceed, I assume that, within the narrow low-energy window of the "soft gap" that drives decoherence, the spatial distribution of the defect states is approximately independent of their specific energy \citep{dynes1978direct}. This allows us to separate the variables in the local density of states: $\rho(x, \epsilon) \approx \rho(x)\gamma(\epsilon)$. Assuming further that the background spatial density $\rho(x)$ is roughly uniform across the highly confined nanoscale dimensions of the junction ($\rho(x) \approx \rho_0$), the macroscopic cross-correlation mathematically factorizes into the product of a purely energy-dependent dynamic integral and a purely spatial overlap integral
\begin{equation}
\propto \left( \int d\epsilon \, \gamma(\epsilon) \langle \nu(\epsilon, t) \eta(\epsilon, 0) \rangle \right) \times \left( \int dx \, \zeta_s(x) \zeta_a(x) \right).
\end{equation}
Because $\zeta_s$ and $\zeta_a$ are distinct eigenstates forming a single fermionic level, their spatial wave functions must be strictly orthogonal. Consequently, the spatial overlap integral exactly vanishes. Therefore, despite the complex time-dependent mixing of the bath Majoranas at the microscopic level, the strict quantum orthogonality of the co-located Majorana modes forces the entire factorized product to evaluate to exactly zero. This macroscopic averaging rigorously guarantees that the dense environment naturally factors into two statistically independent, uncorrelated baths, formally justifying the two-bath spin-boson model in Eq.~(\ref{eq:Q_frustration}).

This mathematical decoupling can be intuitively understood through the lens of standard quantum impurity models, such as the Kondo problem. The crucial bridge between the formal integration above and this physical picture is the separability of the local density of states, $\rho(x, \epsilon) \approx \rho(x)\gamma(\epsilon)$. Because the spatial distribution of the defect states remains consistent across the narrow low-energy window that drives decoherence, and because these low-energy environmental states have wavelengths much longer than the short junction ($L \ll \lambda_{env}$), the co-located Majoranas act as highly localized, point-like scattering centers at the origin of the junction.

Following standard scattering theory, the environmental continuum interacting with the junction can be expanded into distinct scattering channels based on parity. The symmetric Majorana mode, possessing an antinode at the junction center, acts analogously to an $s$-wave scatterer and couples exclusively to the even-parity channel of the bath. Conversely, the antisymmetric Majorana mode, possessing a node at the center, acts analogously to a $p$-wave scatterer and couples exclusively to the odd-parity channel of the bath. Because the even and odd scattering channels of the free-fermion environment are orthogonal by symmetry, they constitute completely independent thermodynamic baths. Thus, the distinct spatial parities of the Majorana wave functions naturally partition the local environment, ensuring the statistical independence of the noise required for frustration of decoherence.

\end{document}